\documentclass[10pt,aps,prl,twocolumn,amssymb,showpacs,preprintnumbers,nofootinbib]{revtex4-1}
\usepackage{amsmath,amssymb,graphicx}
\usepackage[stable]{footmisc}
\usepackage{color}

\pdfoutput=1

\newcommand{\nc}{\newcommand}
\nc{\postscript}[2]
{\setlength{\epsfxsize}{#2\hsize}\centerline{\epsfbox{#1}}}
\nc{\non}{\nonumber}
\nc{\hc}{\hbox {h.c.}} \nc{\re}{\hbox {Re}} 
\nc{\mev}{\hbox {MeV}} \nc{\gev}{\;\hbox {GeV}} \nc{\tev}{\;\hbox {TeV}}
\def\lsim{\mathrel{\raise.3ex\hbox{$<$\kern-.75em\lower1ex\hbox{$\sim$}}}}
\def\gsim{\mathrel{\raise.3ex\hbox{$>$\kern-.75em\lower1ex\hbox{$\sim$}}}}

\nc{\etal}{{\it et al.}}
\nc{\Lsp}{\;\;\;\;\;\;\;\;\;\;}  \nc{\LLLsp}{\lspace \lspace}
\nc{\lsp}{\;\;\;\;\;\;}
\nc{\spac}{\;\;\;}
\nc{\noi}{\noindent}
\nc{\beq}{\begin{equation}}   \nc{\eeq}{\end{equation}}
\nc{\bea}{\begin{eqnarray}}   \nc{\eea}{\end{eqnarray}}
\nc{\baa}{\begin{array}}      \nc{\eaa}{\end{array}}
\nc{\bit}{\begin{itemize}}    \nc{\eit}{\end{itemize}}
\nc{\ben}{\begin{enumerate}}  \nc{\een}{\end{enumerate}}
\nc{\bce}{\begin{center}}     \nc{\ece}{\end{center}}

\def\sq2{\sqrt{2}}

\def\ph{\varphi}

\def\m4{m^4(\ph)}
\def\mn2{m_n^2}

\def\v5{V^{(5)}}

\def\baa{\begin{array}}
\def\eaa{\end{array}}

\begin{document}


\preprint{IZTECH/PHYS-2014-04}

\title{\bf Higgsed Stueckelberg Vector and Higgs Quadratic Divergence}

\vskip 20pt

\author{Durmu\c{s} Ali Demir\footnote{demir@physics.iztech.edu.tr}}
\author{Canan Nurhan Karahan \footnote{cananduzturk@iyte.edu.tr}}
\author{Beste Korutlu\footnote{bestekorutlu@iyte.edu.tr}}
\affiliation{Department of Physics, \.{I}zmir Institute of Technology\\
Urla, \.{I}zmir, 35430 TURKEY}
\date{\today}

\begin{abstract}
Here we show that, a hidden vector field whose gauge invariance is ensured by a Stueckelberg scalar and whose mass is spontaneously generated by the Standard Model Higgs field
contributes to quadratic divergences in the Higgs boson mass squared, and even leads to its cancellation at one-loop when Higgs coupling to gauge field is fine-tuned. In contrast
to mechanisms based on hidden scalars where a complete cancellation cannot be achieved, stabilization here is complete in that the hidden vector and the accompanying Stueckelberg
scalar are both free from quadratic divergences at one-loop. This stability, deriving from hidden exact gauge invariance, can have important implications for modelling dark
phenomena like dark matter, dark energy, dark photon and neutrino masses. The hidden fields can be produced at the LHC.
\end{abstract}


\maketitle

\section{Introduction}\label{sec:introduction}

With the discovery of a new resonance at the Large Hadron Collider (LHC), having a mass $m_h=125.9\pm 0.4$ GeV \cite{Aad:2012tfa} and couplings well consistent with the Standard Model (SM) predictions \cite{couplings}, the Higgs naturalness problem \cite{natural} has become the foremost problem to be tackled. The resolution, if any, brings its own new physics structure.
The squared-masses of fundamental scalars, contrary to chiral fermions and gauge bosons whose masses are protected by chiral and gauge invariances, receive additive quantum corrections proportional to
$\Lambda^2$ -- the UV boundary of the SM. In explicit terms, one-loop quantum correction to Higgs squared-mass, originally computed by Veltman \cite{Veltman:1980mj}, reads as
\bea
\label{VC}
(\delta m_H^2)_{\rm \tiny{quad}}=\frac{\Lambda^2}{16 \pi^2}\left(6\lambda_H+\frac{9}{4}g^2+\frac{3}{4}g'^2-6g_t^2\right),
\eea
where $g$ and $g'$ are the $SU(2)_L$ and $U(1)_Y$ gauge couplings of the SM, respectively, and $g_t=m_t/\upsilon_H$ ($\upsilon_H=246$ GeV is the VEV of the Higgs field) is the top quark Yukawa coupling. The top quark, being the most strongly coupled SM particle to the Higgs field, induces the biggest contribution and ensures a nonvanishing, unremovable coefficient before $\Lambda^2$. The Higgs boson mass is stabilized to electroweak scale if $|\delta m_H^2|<m_H^2<\Lambda^2$. This is the Veltman condition (VC). The parameters in it have all been measured, and it violates the LHC results for
$\Lambda>500$ GeV \cite{natural-np}.

Having no symmetry to prevent the Higgs boson mass from sliding to the higher scales via (\ref{VC}), frequently a cancellation mechanism is implemented via fine-tuning of counter terms in which low and high energy degrees of freedom are mixed. This renders the whole procedure unnatural. It would be more natural, if the cancellation occurs by means of a symmetry principle at higher scales, or if it arises by accidental cancellations of certain terms. In fact, models of new physics constructed to complete the SM beyond Fermi energies have all been motivated by Higgs naturalness problem
\cite{natural-np} (see also \cite{natural-susy} for studies within supersymmetry). So far, however, in the 7 TeV and 8 TeV LHC searches reaching out beyond the {\rm TeV} domain, no compelling sign of evidence for new physics has been found \cite{bsm}.

In consequence, having no TeV scale new physics for achieving naturalness, one is forced to understand the electroweak unnaturalness within the SM plus general relativity, albeit with some imperative
extensions required by specifics of the approach taken. In 1995, conformal symmetry \cite{Bardeen:1995kv} was proposed as a mechanism for solving the Higgs mass hierarchy problem (the latest studies on the conformal symmetry as a solution to the fine-tuning problem may be found in \cite{Kawamura:2013kua}). Recently, the Higgs coupling to spacetime curvature has been found to stabilize the electroweak scale by a harmless, soft fine-tuning \cite{Demir:2014gca}. Furthermore, anti-gravity effects have been claimed to improve Higgs naturalness \cite{agravity}. Alternatively, one may view the parameters chosen by nature as the necessity of existence, and this leads to anthropic considerations \citep{Agrawal:1997gf}. In variance with all these approaches, a fine-tuning method based on singlet scalars
\cite{Bjorken:1991xr} has also been employed. In this approach, main idea is to cancel the quadratic divergences in Higgs boson mass with the loops of the singlet scalars that couple to Higgs field
\cite{fine-tune-scalars}. This method, though a fine-tuning operation by itself, nullifies the quadratic divergences and accommodates viable dark matter candidates \cite{Chakraborty:2012rb, Barger:2007im}.
Nevertheless, for real singlet scalars with vacuum expectation value (VEV), it is not possible to kill the quadratic divergences consistently because there is a mixing between the CP-even component of the Higgs field and the real singlet scalar, and it does not allow for simultaneous cancellation of the quadratic divergences in  Higgs boson and singlet scalar masses \cite{Karahan:2014ola}. There are also studies on two-Higgs doublet models without flavor changing neutral currents, demonstrating that, although the cancellation in the coefficient of the one-loop quadratically divergent terms is possible, the parameter space is severely constrained \cite{Chakraborty:2014oma}. An additional complex scalar triplet extension of the SM has also been studied and proven to be a solution to the fine-tuning problem
\cite{Chakraborty:2014xqa}.

In the present work, as a completely new approach never explored before, we study protection of the Higgs boson mass by a SM-singlet gauge field (not a scalar field as in \cite{fine-tune-scalars}). In contrast to the attempts based on hidden scalars \cite{fine-tune-scalars, Chakraborty:2012rb,Chakraborty:2014oma,Chakraborty:2014xqa}, which are now known to be unable to simultaneously protect the masses of the Higgs boson and the singlet scalar \cite{Karahan:2014ola}, in the present work, we consider a hidden $U(1)$ gauge field $V_{\mu}$ whose invariance is ensured by a Stueckelberg scalar $S$ and whose mass is spontaneously induced by the SM Higgs field. We show that $V_{\mu}$ and $S$ enable cancellation of the quadratic divergence in Higgs boson mass with no quadratic divergence arising in their own masses. It is important that the SM Higgs boson is stabilized at one-loop along with already-stable hidden gauge and Stueckelberg scalar. This phenomenological advantage has important implications not only for stabilizing the Higgs boson mass but also for correlating the SM Higgs field with hidden sectors.

The paper is organized as follows. In Section 2 below, we construct the model starting from the basic Stueckelberg setup. Section 3 is devoted to computation of the quadratic divergences and vanishing of the Higgs mass divergence by fine-tuning. We conclude in Section 4.

\section{The Model}
In this section, we consider a massive Abelian gauge field $V_{\mu}$ accompanied by a real scalar field $S(x)$, introduced to preserve the gauge invariance of the theory. Originally proposed by Stueckelberg \cite{Stueckelberg:1938zz} and noted afterwards by Pauli \cite{Pauli:1941zz} that, $V_{\mu}$ satisfies a restricted $U(1)$ gauge invariance, with the gauge function $\Theta(x)$ obeying a massive Klein-Gordon equation. The mechanism provides an alternative to the Higgs mechanism, where the vector boson acquires its mass with the breakdown of the gauge invariance of not the Lagrangian but of the vacuum. These features are encoded in the Stueckelberg model \cite{nath}
\bea
\label{basic}
\mathcal{L}=\!-\frac{1}{4}V_{\mu\nu}^2\!+\frac{1}{2}m^2\!\!\left(\!V^{\mu}\!\!-\frac{1}{m}\partial^{\mu}S\!\right)^2\!\!\!\!
\!-\!\frac{1}{2}(\partial_{\mu}V^{\mu}\!+m S)^2\!,
\eea
where $m$ is the common mass for $V_{\mu}$ and $S$. Despite its massive spectrum, this model enjoys a $U(1)_{m}$ invariance
\bea
\label{U(1)m}
&&V_{\mu}(x)\rightarrow V'_{\mu}(x)=V_{\mu}(x)+\partial_{\mu}\Theta(x),\non\\
&&S(x)\rightarrow S'(x)=S(x)+m\Theta(x)\!,
\eea
provided that $\left( \Box + m^2 \right) \Theta(x) = 0$. Consequently, in spite of its nonvanishig hard mass, $V_{\mu}$
enjoys exact gauge invariance, albeit with a restricted gauge transformation function $\Theta(x)$ \cite{nath}. In the massless
limit, $m \rightarrow 0$, the Stueckelberg Lagrangian (\ref{basic}) reduces to $\mathcal{L}_{m=0} = -\frac{1}{4}V_{\mu\nu}^2 +
\frac{1}{2} \partial_{\mu}S\, \partial^{\mu}S$, which is obviously $U(1)_{m}$ invariant in Lorentz gauge ($\partial_{\mu}V^{\mu}=0$) with an unrestricted $\Theta(x)$.
Interestingly, the Stueckelberg scalar $S$, transforming like the gauge field $V_{\mu}$ in massive case, turns into a
gauge-singlet scalar in massless limit.

Inspired from the Stueckelberg model (\ref{basic}), we propose the Higgsed Stueckelberg model
\bea
\label{HS}
\mathcal{L}=&-&\frac{1}{4}V_{\mu\nu}^2+\lambda_1 H^{\dagger}H\left(V^{\mu}-\frac{1}{\sqrt{\lambda_1}a_H}\partial^{\mu}S\right)^2\non\\
&-&\frac{1}{2}(\partial_{\mu}V^{\mu}+\sqrt{\lambda_1}a_H S)^2,
\eea
where $\lambda_1$ is a positive dimensionless constant and $a_H$ is a mass parameter. This model is manifestly gauge-invariant
under both the hidden $U(1)_m$ invariance with $m\rightarrow \sqrt{\lambda_1}a_H$, and the electroweak gauge group $SU(2)_L\otimes U(1)_Y$. The Higgs potential
$V(H) = m_H^2 H^{\dagger} H + \lambda_H \left(H^{\dagger} H \right)^{2}$ and hence the total energy is minimized at the
Higgs field configuration
\begin{eqnarray}
\label{higgs-VEV}
\langle H^{\dagger} H \rangle = \begin{cases}
\frac{\upsilon_{H}^2}{2} & \text{if}\,\,\,\, m_H^2<0,\\
0 & \text{if}\,\,\,\, m_H^2>0,
\end{cases}
\end{eqnarray}
where $\upsilon_{H}= \sqrt{-\frac{m_H^2}{\lambda_H}}$ is the Higgs VEV in the broken
phase ($m_H^2<0$), to which masses of the SM particles are all proportional. In this phase electroweak gauge group
$SU(2)_{L}\otimes U(1)_Y$ is spontaneously broken down to electromagnetism. In unbroken phase ($m_H^2>0$) electroweak
group stays exact and all the SM particles but Higgs boson are massless.

From (\ref{HS}) it is clear that, the two phases of the SM directly leave distinguishable effects on the mass of $V_{\mu}$ and
kinetic term of $S$. And the Stueckelberg structure in (\ref{basic}) is achieved properly if the mass parameter $a_H$
can keep track of the two electroweak phases. This feature is implemented into the Higgsed Stueckelberg model (\ref{HS}) by setting
\bea
\label{aH}
a_H=\Re \left(\sqrt{-\frac{m_H^2}{\lambda_H}}\right)=
\begin{cases} \upsilon_H & \text{if}\,\,\,\, m_H^2<0,\\
0 & \text{if}\,\,\,\, m_H^2>0,
\end{cases}
\eea
which obviously dogs the Higgs VEV in (\ref{higgs-VEV}). It turns out that $\langle H^{\dagger} H \rangle  = a_H^2/2$ in both broken and exact electroweak phases, and
$\upsilon_H = a_H$ specifically in the broken phase. This switching ability of $a_H$ ensures that, in the broken phase of electroweak group, there arises, in addition to
the massive SM spectrum, a massive vector $V_{\mu}$ with mass $M_V^2 = \lambda_1 \upsilon_H^2$ and a massive scalar $m_S^2 =\lambda_1  a_H^2$. In the unbroken phase, however,
the Higgs field stands as the only massive field. The rest, inclusing $V_{\mu}$ and $S$, are all massless. In what follows, we will work in the physical vacuum of the
broken electroweak phase and necessarily set $a_H = \upsilon_H$ everywhere.

It is instructive to study the transcription of the Stueckelberg $U(1)_{m}$ symmetry in (\ref{U(1)m}) into the Higgsed Stueckelberg case. To this end, one notes that the Stueckelberg scalar $S(x)$ facilitates $U(1)_m$ gauge invariance of the hidden sector, and also, helps keep the Hamiltonian positive definite\footnote{Note that the last term in (\ref{HS}) can also be written as $\mathcal{L}_{\rm gf}=-\frac{1}{2\alpha}(\partial_{\mu}V^{\mu}+\alpha\sqrt{\lambda_1}\upsilon_H S)^2$, where $\alpha$ is a real parameter, similar to t'Hooft's parametrization for Abelian Higgs model. The choice of $\alpha=1$ corresponds to the Stueckelberg-Feynman gauge. When $\alpha\neq 1$, the restriction on the gauge function changes to $(\Box+\alpha\lambda_1\upsilon_H^2)\Lambda(x)=0$. It is also possible to choose two different parameters $\alpha_1$ and $\alpha_2$, to check the gauge independence of the parameters. However, there is the disadvantage that the terms of the form $V^{\mu}\partial_{\mu}B$ survives for this choice. In the present work, we will work in Stueckelberg-Feynman gauge.} \cite{Stueckelberg:1938zz}. In this formalism, Lorentz subsidiary condition does not follow from equation of motion. Imposing an operator equation of the form $\partial^{\mu}V_{\mu}^{(-)}(x)|{\rm\bf phys\rangle}=0$, where $V_{\mu}^{(-)}(x)$ involves the free field annihilation operators, however, gives rise to conflict between the operator equation and  the canonical commutation relations. This puzzle is solved via the introduction of an additional scalar field $S(x)$, replacing the operator equation with $\Phi(x)|{\rm\bf phys\rangle}\equiv[\partial^{\mu}V_{\mu}^{(-)}(x)+m S^{(-)}(x)]|{\rm\bf phys\rangle}=0$, where $S^{(-)}(x)$ also involves free field annihilation operators. The operator equation decreases the number of degrees of freedom of the Lagrangian to four. The required constraint to decrease it to three for a massive vector field comes into play with the gauge transformation
\bea
\label{U(1)mH}
&&V_{\mu}(x)\rightarrow V'_{\mu}(x)=V_{\mu}(x)+\partial_{\mu}\Theta(x),\non\\
&&S(x)\rightarrow S'(x)=S(x)+\sqrt{\lambda_1} \upsilon_H \Theta(x)\!,
\eea
which closely follows the Stueckelberg transformation (\ref{U(1)m}). The $U(1)_m$ invariance is ensured if $(\partial^2+\lambda_1 \upsilon_H^2)\Theta(x)=0$. This restricted gauge invariance changes to an
unrestricted, standard gauge invariance in the unbroken ($m_H^2>0$) electroweak phase in which $V_{\mu}$ and $S$ are massless and non-interacting. Moreover, $S$ is a gauge singlet in this phase. The $V_{\mu}$ and its Stueckelberg companion $S$ do possess identical masses in broken and unbroken phases of the electroweak symmetry. In broken phase, Stueckelberg-Feynman gauge, their propagators read as
\bea
\Delta_{\mu\nu}=-\frac{i\, g_{\mu\nu}}{q^2-m^2},\qquad \Delta=\frac{i}{q^{2}-m^{2}},
\eea
where $m^2=\lambda_1\upsilon_H^2$ is the common mass for $V_{\mu}$ and $S$.

\section{Phenomenology}\label{sec:pheno}
\begin{figure}[h]
  \centering
  \includegraphics[scale=0.69]{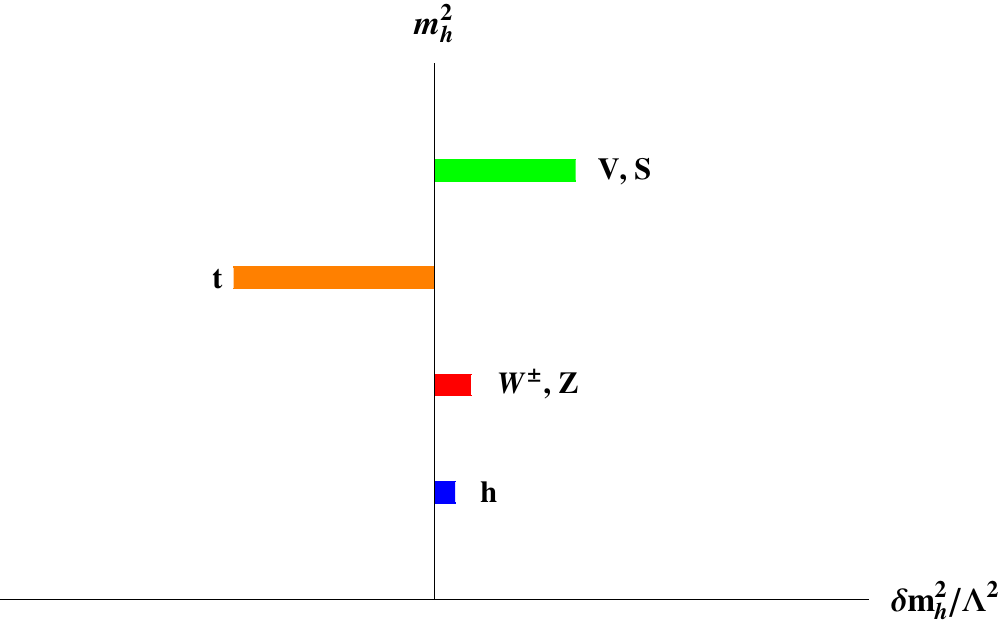}
  \caption{The schematic representation of the quadratically divergent contributions to Higgs boson mass at one-loop level. Here, $h$ denotes the Higgs boson, $W^{\pm}, Z$ the electroweak bosons, $t$ the top quark, $V,S$ the hidden gauge boson $V_{\mu}$ and the Stueckelberg scalar $S$, respectively. Higgs mass is protected from destabilizing quantum effects when the hidden gauge sector is included.} \label{fig:BC1}
\end{figure}

In this section we study quantum corrections to masses of the Higgs boson $h$ and Stueckelberg fields $S$ and $V_{\mu}$. The main constraint on the model is that Higgs boson must weigh $m_h=125.9$ GeV \cite{Aad:2012tfa}. As follows from (\ref{HS}), there are three-point and four-point interactions among the vector boson $V_{\mu}$, the Stueckelberg field $S$, and the Higgs field $h$. The vertex factors are summarized in the Appendix. The Higgsed Stueckelberg hidden sector then modifies the Veltman condition (\ref{VC}) as
\bea
(\delta m_H^2)_{\rm\tiny{quad}}\!=\!\frac{\Lambda^2}{16 \pi^2}\!\left(\!\frac{11}{3}\lambda_H\!+\frac{9}{4}g^2\!+\frac{3}{4}g'^2\!-6g_t^2\!+\lambda_1\!\right)\!,
\eea
wherein $\lambda_1$ shows up as a new degree of freedom. In the philoshopy of the original attempts in \cite{fine-tune-scalars}, one can suppress $(\delta m_H^2)_{\rm\tiny{quad}}$ by choosing $\lambda_1$ appropriately. In particular, $(\delta m_H^2)_{\rm\tiny{quad}}$ vanishes for $\lambda_1=4.41$. The $V_{\mu}$ and $S$ are degenerate in mass, and for this specific value of $\lambda_1$  they weigh  $m=\sqrt{\lambda_1}\upsilon_H=517$ GeV. It is possible to decrease the value of $\lambda_1$ by simply introducing $N$ such fields, which in turn lowers the masses of the new fields while increasing their number. In Figure \ref{fig:BC1}, a schematic representation of the one-loop quantum corrections to Higgs mass is shown in our extended scenario. As it is apparent from this figure, a hidden Abelian gauge sector splendidly cancels the quadratically divergent contributions to Higgs mass from the SM fields.

It is clear that suppressing $(\delta m_H^2)_{\rm\tiny{quad}}$ requires $\lambda_1$ to be finely tuned. The fine-tuning here is of the same size as the fine-tunings required for hidden
scalar sectors \cite{fine-tune-scalars,Chakraborty:2012rb, Karahan:2014ola,Chakraborty:2014oma,Chakraborty:2014xqa}. There is one big difference, however. Indeed, these models based on hidden
scalars suffer from the fact that masses of the hidden scalars and of the SM Higgs boson cannot be protected simultaneously \cite{Karahan:2014ola}. The hidden scalar continues to have a mass ${\mathcal{O}}(\Lambda)$ after suppressing the radiative contribution to the Higgs boson mass. In the Higgs-Stueckelberg model this impasse is overcome. To see this, one notes that mass of the Stueckelberg field does actually receive quadratically divergent radiative corrections from two self energy diagrams (one with Higgs boson in the loop and another with both Higgs and the Stueckelberg field $S$ in the loop). The self energy diagram with a Higgs boson and vector boson $V_{\mu}$ in the loop diverges logarithmically. The spruceness of this scenario emerges at this point in that the quadratically-divergent contributions to the mass of the Stueckelberg field from the two loop diagrams cancel out to give
\bea
(\delta m_S^2)_{\rm \tiny{quad}}=0.
\eea
In the same manner, the mass of $V_{\mu}$ is protected against quadratically-divergent quantum corrections
\bea
(\delta m_V^2)_{\rm \tiny{quad}}=0.
\eea
Leaving aside the logarithmic corrections, masses of $V{\mu}$ and $S$ are found to be UV-insensitive. This is actually expected by gauge invariance because there exists an unbroken $U(1)_m$ invariance in both broken and unbroken electroweak phases. The invariance protects the mass of $V_{\mu}$. Interestingly, it also protects the mass of $S$ because $S$ by itself acts like a gauge field when $V_{\mu}$ is massive and becomes a non-interacting $U(1)_m$ singlet when $V_{\mu}$ is massless. Clearly, the radiative stability of the hidden sector can have important implications for modelling `dark phenomena' like Dark Matter, Dark Energy, Dark Photon and neutrino masses.

\section{Conclusion and Outlook}

The discovery of a new scalar  \cite{Aad:2012tfa} at the LHC, consistent with the SM Higgs boson, has accelerated studies on the UV-sensitivity of the Higgs boson. As opposed to the physical masses of chiral fermions and gauge bosons, which are protected by chiral and gauge symmetries, there is no symmetry principle to protect the Higgs boson mass against quadratically divergent quantum corrections.
In the very absence of TeV-scale new physics, one is left with a finely-tuned Higgs sector where nature and degree of fine-tuning vary with the modeling details. In the presence of hidden scalars, despite the protection of the Higgs boson mass the hidden sector itself is UV-unstable. In case the hidden sector is formed by the spacetime curvature scalar, the fine-tuning is severe yet harmless because the SM fields and couplings are immune to its presence. The fine-tuning is as severe as hidden scalars in other field-theoretic approaches.

In this Letter we have shown that a hidden sector spanned by an Abelian vector field whose mass is induced by electroweak breaking and whose gauge invariance is sustained by a Stueckelberg scalar
can lead to stabilization of the Higgs boson mass by finely tuning its coupling to the SM Higgs field. In spite of this unavoidable fine-tuning, the Higgsed Stueckelberg model possesses
the striking property that the hidden sector is insensitive to the UV scale. This stability, deriving from unbroken hidden gauge invariance, can have important collider, astrophysical and cosmological implications. Indeed, a stable hidden sector can be utilized in constructing viable models of Dark Matter, Dark Energy, Dark Photon and neutrino masses. The model can be tested at the LHC (and its
successor FCC) via direct productions of $V_{\mu}$ and $S$ fields.

\section{Appendix}
Here we list the vertex factors:
\begin{eqnarray}
&&\lambda_{\rm hhVV}=\!2i\lambda_{1}g^{\mu \nu},\non\\
&&\lambda_{\rm hhSS}=\!-\frac{2i}{\upsilon_{H}^2}k_{\mu}q_{\nu}g^{\mu \nu},\non\\
&&\lambda_{\rm hVV}=\!2i\lambda_{1}\upsilon_{H}g^{\mu \nu},\non\\
&&\lambda_{\rm hSS}=-\frac{2i}{\upsilon_{H}}k_{\mu}q_{\nu}g^{\mu \nu},\non\\
&&\lambda_{\rm hVS}=\!2\sqrt{\lambda_{1}}k_{\mu}g^{\mu \nu},
\end{eqnarray}
where $k_{\mu}$ is the momentum of $S$. We used $a_H = \upsilon_H$ in $a_H$ dependent vertices.

\bigskip
\noindent\textbf{Acknowledgments}\\
This work has been supported in part by T\"{U}B\.{I}TAK, The Scientific and Technical Research Council of Turkey, through the grant 2232, Project No: 113C002 and by TAEK, Turkish Atomic Energy Authority, Project No: CERN-A5.H2.P1.01-21. 
%


\begin{thebibliography}{9}
\bibitem{Aad:2012tfa}
  G.~Aad {\it et al.}  [ATLAS Collaboration],
  Phys.\ Lett.\ B {\bf 716}, 1 (2012)
  [arXiv:1207.7214 [hep-ex]];
%
  S.~Chatrchyan {\it et al.}  [CMS Collaboration],
  Phys.\ Lett.\ B {\bf 716}, 30 (2012)
  [arXiv:1207.7235 [hep-ex]].
%
\bibitem{couplings}
 J.~Ellis and T.~You,
  JHEP {\bf 1306}, 103 (2013)
  [arXiv:1303.3879 [hep-ph]];
 A.~Djouadi and G.~ég.~Moreau,
  arXiv:1303.6591 [hep-ph].

\bibitem{natural}
V.~F.~Weisskopf,
  Phys.\ Rev.\  {\bf 56}, 72 (1939);
 K.~G.~Wilson,
  Phys.\ Rev.\ D {\bf 3}, 1818 (1971);
L.~Susskind,
  Phys.\ Rev.\ D {\bf 20}, 2619 (1979).

%
\bibitem{Veltman:1980mj}
  M.~J.~G.~Veltman,
  Acta Phys.\ Polon.\ B {\bf 12}, 437 (1981).
%

\bibitem{natural-np}
J.~L.~Feng,
  arXiv:1302.6587 [hep-ph];
J.~D.~Wells,
  arXiv:1305.3434 [hep-ph];
G.~F.~Giudice,
  arXiv:1307.7879 [hep-ph];
G.~Altarelli,
  Phys.\ Scripta T {\bf 158}, 014011 (2013)
  [arXiv:1308.0545 [hep-ph]].

\bibitem{natural-susy}
  M.~Liu and P.~Nath,
  Phys.\ Rev.\ D {\bf 87}, no. 9, 095012 (2013)
  [arXiv:1303.7472 [hep-ph]];
%
  I.~Masina and M.~Quiros,
  Phys.\ Rev.\ D {\bf 88}, 093003 (2013)
  [arXiv:1308.1242 [hep-ph]];
%
  X.~Lu, H.~Murayama, J.~T.~Ruderman and K.~Tobioka,
  arXiv:1308.0792 [hep-ph];
%
  A.~Fowlie,
  arXiv:1403.3407 [hep-ph].

\bibitem{bsm}
M.~Flechl [CMS and ATLAS Collaborations],
  arXiv:1307.4589 [hep-ex];
  J.~L.~Feng, J.~-F.~Grivaz and J.~Nachtman,
  Rev.\ Mod.\ Phys.\  {\bf 82}, 699 (2010)
  [arXiv:0903.0046 [hep-ex]].
%
\bibitem{Bardeen:1995kv}
  W.~A.~Bardeen,
  FERMILAB-CONF-95-391-T.
%
\bibitem{Kawamura:2013kua}
  D.~A.~Demir,
  arXiv:1207.4584 [hep-ph];
  %
  M.~Heikinheimo, A.~Racioppi, M.~Raidal, C.~Spethmann and K.~Tuominen,
  arXiv:1304.7006 [hep-ph];
%
  G.~M. Tavares, M.~Schmaltz and W.~Skiba,
  Phys.\ Rev.\ D {\bf 89}, 015009 (2014)
  [arXiv:1308.0025 [hep-ph]];
%
  Y.~Kawamura,
  PTEP {\bf 2013}, no. 11, 113B04 (2013)
  [arXiv:1308.5069 [hep-ph]];
%
  M.~Holthausen, J.~Kubo, K.~S.~Lim and M.~Lindner,
  JHEP {\bf 1312}, 076 (2013)
  [arXiv:1310.4423 [hep-ph]];
%
  O.~Antipin, M.~Mojaza and F.~Sannino,
  Phys.\ Rev.\ D {\bf 89}, 085015 (2014)
  [arXiv:1310.0957 [hep-ph]];
%
  J.~Guo and Z.~Kang,
  arXiv:1401.5609 [hep-ph].
  %
\bibitem{Demir:2014gca}
  D.~A.~Demir,
  arXiv:1405.0300 [hep-ph].
%
\bibitem{agravity}
A.~Salvio and A.~Strumia,
  JHEP {\bf 1406} (2014) 080
  [arXiv:1403.4226 [hep-ph]].
%
\bibitem{Agrawal:1997gf}
  V.~Agrawal, S.~M.~Barr, J.~F.~Donoghue and D.~Seckel,
  Phys.\ Rev.\ D {\bf 57}, 5480 (1998)
  [hep-ph/9707380].
%
\bibitem{Bjorken:1991xr}
  R.~S.~Chivukula and M.~Golden,
  Phys.\ Lett.\ B {\bf 267}, 233 (1991);
%
  R.~S.~Chivukula, M.~Golden and M.~V.~Ramana,
  Phys.\ Lett.\ B {\bf 293}, 400 (1992)
  [hep-ph/9206255];
%
  J.~D.~Bjorken,
  Int.\ J.\ Mod.\ Phys.\ A {\bf 7}, 4189 (1992);
%
\bibitem{fine-tune-scalars}
M. Ruiz-Altaba, B. González, M. Vargas, CERN-TH.5558/89 (1989);
M.~Capdequi Peyranere, J.~C.~Montero and G.~Moultaka,
  Phys.\ Lett.\ B {\bf 260}, 138 (1991);
A.~A.~Andrianov, R.~Rodenberg and N.~V.~Romanenko,
  Nuovo Cim.\ A {\bf 108}, 577 (1995)
  [hep-ph/9408301];
A.~Kundu and S.~Raychaudhuri,
  Phys.\ Rev.\ D {\bf 53} (1996) 4042
  [hep-ph/9410291];
 F.~Bazzocchi, M.~Fabbrichesi and P.~Ullio,
  Phys.\ Rev.\ D {\bf 75} (2007) 056004
  [hep-ph/0612280].
\bibitem{Chakraborty:2012rb}
  I.~Chakraborty and A.~Kundu,
  Phys.\ Rev.\ D {\bf 87}, no. 5, 055015 (2013)
  [arXiv:1212.0394 [hep-ph]].
%
\bibitem{Barger:2007im}
  J.~McDonald,
  Phys.\ Rev.\ D {\bf 50}, 3637 (1994)
  [hep-ph/0702143 [HEP-PH]];
%
  D.~A.~Demir,
  Phys.\ Lett.\ B {\bf 450}, 215 (1999)
  [hep-ph/9810453];
%
  C.~P.~Burgess, M.~Pospelov and T.~ter Veldhuis,
  Nucl.\ Phys.\ B {\bf 619}, 709 (2001)
  [hep-ph/0011335];
%
  V.~Barger, P.~Langacker, M.~McCaskey, M.~J.~Ramsey-Musolf and G.~Shaughnessy,
  Phys.\ Rev.\ D {\bf 77}, 035005 (2008)
  [arXiv:0706.4311 [hep-ph]];
%
  V.~Barger, P.~Langacker, M.~McCaskey, M.~Ramsey-Musolf and G.~Shaughnessy,
  Phys.\ Rev.\ D {\bf 79}, 015018 (2009)
  [arXiv:0811.0393 [hep-ph]];
%
  W.~-L.~Guo and Y.~-L.~Wu,
  JHEP {\bf 1010}, 083 (2010)
  [arXiv:1006.2518 [hep-ph]];
%
  A.~Djouadi, O.~Lebedev, Y.~Mambrini and J.~Quevillon,
  Phys.\ Lett.\ B {\bf 709}, 65 (2012)
  [arXiv:1112.3299 [hep-ph]];
%
  M.~Gonderinger, H.~Lim and M.~J.~Ramsey-Musolf,
  Phys.\ Rev.\ D {\bf 86}, 043511 (2012)
  [arXiv:1202.1316 [hep-ph]];
%
  B.~Batell, D.~McKeen and M.~Pospelov,
  JHEP {\bf 1210}, 104 (2012)
  [arXiv:1207.6252 [hep-ph]];
%
  S.~Baek, P.~Ko, W.~-I.~Park and E.~Senaha,
  JHEP {\bf 1211}, 116 (2012)
  [arXiv:1209.4163 [hep-ph]];
%
  A.~Biswas and D.~Majumdar,
  Pramana {\bf 80}, 539 (2013)
  [arXiv:1102.3024 [hep-ph]];
%
  G.~Belanger, K.~Kannike, A.~Pukhov and M.~Raidal,
  JCAP {\bf 1301}, 022 (2013)
  [arXiv:1211.1014 [hep-ph]];
%
  J.~M.~Cline and K.~Kainulainen,
  JCAP {\bf 1301}, 012 (2013)
  [arXiv:1210.4196 [hep-ph]];
%
  D.~A.~Demir, M.~Frank and B.~Korutlu,
  Phys.\ Lett.\ B {\bf 728}, 393 (2014)
  [arXiv:1308.1203 [hep-ph]];
%
  N.~Haba, K.~Kaneta and R.~Takahashi,
  JHEP {\bf 1404}, 029 (2014)
  [arXiv:1312.2089 [hep-ph]].
%
\bibitem{Karahan:2014ola}
  C.~N.~Karahan and B.~Korutlu,
   Phys.\ Lett.\ B {\bf 732}, 320 (2014)
 [arXiv:1404.0175 [hep-ph]].
%
\bibitem{Chakraborty:2014oma}
  I.~Chakraborty and A.~Kundu,
  arXiv:1404.3038 [hep-ph].
%
\bibitem{Chakraborty:2014xqa}
  I.~Chakraborty and A.~Kundu,
  arXiv:1404.1723 [hep-ph].


\bibitem{Stueckelberg:1938zz}
  E.~C.~G.~Stueckelberg,
  Helv.\ Phys.\ Acta {\bf 11}, 299 (1938);
  Helv.\ Phys.\ Acta {\bf 11}, 225 (1938).
%
\bibitem{Pauli:1941zz}
  W.~Pauli,
  Rev.\ Mod.\ Phys.\  {\bf 13}, 203 (1941).
%
\bibitem{nath}
  B.~Kors and P.~Nath,
  Phys.\ Lett.\ B {\bf 586}, 366 (2004)
  [hep-ph/0402047];
  JHEP {\bf 0507}, 069 (2005)
  [hep-ph/0503208].
\end{thebibliography}
\end{document}